\documentclass[conference]{IEEEtran}
\IEEEoverridecommandlockouts
\usepackage{cite}
\usepackage{amsmath,amssymb,amsfonts}
\usepackage{algorithmic}
\usepackage{booktabs,threeparttable}
\usepackage{graphicx}
\usepackage{tablefootnote}
\usepackage{hyperref}
\usepackage{textcomp}
\usepackage{xcolor}
\usepackage{verbatim}

\def\BibTeX{{\rm B\kern-.05em{\sc i\kern-.025em b}\kern-.08em
    T\kern-.1667em\lower.7ex\hbox{E}\kern-.125emX}}
\begin{document}

\title{Performance landscape of resource-constrained platforms targeting DNNs
}

\author{\IEEEauthorblockN{Panagiotis Miliadis}
\IEEEauthorblockA{\textit{National Technical University of Athens} \\
pmiliad@cslab.ece.ntua.gr}
\and
\IEEEauthorblockN{Christos-Savvas
Bouganis}
\IEEEauthorblockA{\textit{Imperial College London}\\
christos-savvas.bouganis@imperial.ac.uk}
\and
\IEEEauthorblockN{Dionisios Pnevmatikatos}
\IEEEauthorblockA{\textit{National Technical University of Athens}\\
pnevmati@cslab.ece.ntua.gr}
}

\maketitle

\thispagestyle{plain}
\pagestyle{plain}

\begin{abstract}
Over the recent years, a significant number of complex, deep neural networks have been developed for a variety of applications including speech and face recognition, computer vision in the areas of health-care, automatic translation, image classification, etc. Moreover, there is an increasing demand in deploying these networks in resource-constrained edge devices. 
As the computational demands of these models keep increasing, pushing to their limits the targeted devices, the constant development of new hardware systems tailored to those workloads has been observed.
Since programmability of these diverse and complex platforms -compounded by the rapid development of new DNN models- is a major challenge, platform vendors have developed Machine Learning tailored SDKs to maximize the platform's performance.

This work investigates the performance achieved on a number of modern commodity embedded platforms coupled with the vendors' provided software support when state-of-the-art DNN models from image classification, object detection and image segmentation are targeted. The work quantifies the relative latency gains of the particular embedded platforms and provides insights on the relationship between the required minimum batch size for achieving maximum throughput, concluding that modern embedded systems reach their maximum performance even for modest batch sizes when a modern state of the art DNN model is targeted. Overall, the presented results provide a guide for the expected performance for a number of state-of-the-art DNNs on popular embedded platforms across the image classification, detection and segmentation domains.
\end{abstract}


\section{Introduction}

In the recent years, technological advancements in computational power have fueled an impressive progress in the development of AI tasks based on deep learning. The introduction of AlexNet\cite{AlexNet} in 2012, marked a new beginning in Deep Neural Networks (DNN) and computer vision space. Since then, many models that could challenge -or even outperform- human estimation accuracy have been developed for a variety of computer vision tasks. However, the gains in terms of accuracy of a model, usually come at the cost of longer inference time and high resource requirements.
The execution of DNN tasks quickly finds its way into handheld, embedded, or edge platforms, that are typically resource constrained. The choice of the suitable platform depends mainly on the power consumption of the system and its performance on specific tasks.  This work captures the performance characterisation of a number of widely available resource-constrained embedded devices on three widely adopted inference tasks. The selected models are presented in \autoref{DNNModels}, alongside with information regarding their way of utilization and computational demands.

\begin{table}[t]
\caption{DNN models used in the evaluation}
\vspace{-0.05in}
\resizebox{\columnwidth}{!}{
\setlength\tabcolsep{2.5pt} 
\begin{tabular}{llllrrr}
\hline
\textbf{Models} & \textbf{Task} & \textbf{Dataset} & \textbf{Year} & \textbf{MAC(G)} & \textbf{Params(M)} & \textbf{Top-1}    \\
\textbf{}       & \textbf{}     & \textbf{}        & \textbf{}     & \textbf{}       & \textbf{}          & \textbf{  Accuracy}\\ \hline
Alexnet\cite{AlexNet} & Image & ImageNet & 2012 & 0.73 & 60.97 & 56.62\%\\
InceptionV1\cite{IncV1} & Classification & ImageNet & 2014 & 1.59 & 7.00 & 68.70\%\\
InceptionV4\cite{IncV4} &  & ImageNet & 2017 & 12.27 & 42.62 & 80.08\%\\
SqueezeNetV1.1\cite{squeeze} &  & ImageNet & 2016 & 0.39 & 1.24 & 58.18\%\\
DenseNet121\cite{dense} &  & ImageNet & 2017 & 3.08 & 7.98 & 74.43\%\\
DenseNet161\cite{dense} &   & ImageNet & 2017 & 8.52 & 28.73 & 77.14\%\\
DenseNet169\cite{dense} &  & ImageNet & 2017 & 3.72 & 14.15 & 75.60\%\\
ResNet50\cite{res} &   &ImageNet & 2015 & 3.87 & 25.56 & 76.13\%\\
ResNet101\cite{res} &  & ImageNet & 2015 & 7.59 & 44.55 & 77.37\%\\
ResNet152\cite{res} &  & ImageNet & 2015 & 11.30 & 60.19 & 78.31\%\\
VGG16\cite{vgg} &  & ImageNet & 2014 & 15.47 & 138.36 & 71.59\%\\
VGG19\cite{vgg} &  & ImageNet & 2014 & 19.63 & 143.67 & 72.38\%\\ 
MobileNetV1\cite{MobileNets} &  & ImageNet & 2017 & 0.57 & 4.23 & 70.81\%	\\
MobileNetV2\cite{MobileNets} &  & ImageNet & 2017 & 0.44 & 3.51 & 71.90\%\\ \hline
SSD300\cite{ssd} & Object & VOC0712 & 2016 & 31.37 & 26.28 & 75.8 mAP\\
SSD512\cite{ssd} & Detection & VOC0712 & 2016 & 90.21 & 27.19 & 78.5 mAP\\
YOLO-V3\cite{yolov3} &  & COCO & 2018 & 27.27 & 61.87 & 55.3 mAP\\ \hline
FCN8\cite{fcn8} & Image  & VOC0712 & 2015 & 181.55 & 134.49 & 75.9\%\\
Dilation\cite{dilation} & Segmentation & CityScapes & 2015 & 2650.00 & 134.46 & - \tablefootnote{No official report is available for the specific dataset.} \\ \hline
\end{tabular}}
\label{DNNModels}
\vspace{-0.15in}
\end{table}

A large spectrum of the available platforms are combined with a software development kit (SDK) tailored for the optimised execution of DNNs. These kits provide software support and include a set of tools and libraries that implement low-level functions for the specific hardware architecture, aiming to maximise the system's performance under machine learning workloads. The existence of such software support is instrumental for the hardware platform to deliver high performance under machine learning workloads, dictating the consideration of the available SDK when it comes to the performance comparison of these devices under ML loads.

The focus of this study is on characterizing the performance of widely utilised embedded systems for inference in real-life applications. The paper provides information on the obtained latency of the systems under investigation under batch size of 1, aiming to characterise applications that are latency-critical, as well as larger batch sizes to explores the batch-size impact on  system throughput for throughput-oriented applications. Furthermore, we extend our analysis by providing real-time power consumption measurements across all different inference workloads, as well on energy efficiency of the embedded platforms executing each available model.  

The following sections of the paper are organized as follows. Section II presents related previous works. Section III describes the hardware platforms that were used along with a brief overview of the available toolkits for accelerating the DNN inference. Sections IV and V present the experimental results and a discussion on them respectively, and Section V provides our conclusions.

\section{Related Work}

Some notable works have analyzed the execution performance of DNN models on various deployment platforms. In their work \cite{blott}, Blott {\em et. al.} are focused on how accelerators are benefit from different optimizations techniques, like pruning, quantization and numerical representations. However, their experiments include a small number of available public models, and they focus only on performance capabilities of their platforms, rather than real power consumption and energy efficiency. Ignatov {\em et. al.} in \cite{AI} present an extended benchmark analysis of the Android smartphones' performance, by running various DNNs and tasks on their available System on Chip (SoC). Another approach for increasing the performance of Convolutional Neural Networks (CNNs) was made by Hegde {\em et. al.} in \cite{CaffePresso}. Their work is focused on running CNNs on hardware accelerators which are integrated in commodity SoCs platforms. Almeida {\em et. al.} \cite{Embench} present a deep performance analysis of modern commodity platforms that can be used on desktops and server environments. Their work is mainly focused on running DNN models for the image classification task but without utilizing the available SDK toolsets which might accelerate the inference execution. MLPerf benchmark \cite{mlperf} reports a number of results on both inference and training tasks focusing on specific scenarios and metrics. The authors have provided a protocol for users to investigate the performance of the system and upload their results. 

This work is aligned with the above approaches and contributes to the field by focusing on the performance characterisation of a number of resource-constrained platforms, under a large number of workloads across three difference computer vision tasks, providing insights on their performance as well on their real-time power consumption and energy efficiency. 

\section{Hardware Platforms} \label{HW}

Our performance landscape is based on resource-constrained platforms. These platforms are usually integrated into embedded systems or desktop environment where both computational and memory resources are limited. By following this approach, our platform list includes embedded SoCs that are especially designed for inference tasks, as well general purpose units that are commonly included on many modern commodity systems.

\subsection{Intel}

The spectrum of Intel's platforms is very wide, as it includes numerous processor units for either desktops or high-end server environments. A conventional solution, for resource constrained systems and for edge computing, is a high end general purpose CPU, such as i7 6700. This processor unit offers 4 physical cores, but with HyperThreading enabled it reaches 8 logical cores. Also, its base frequency is set at 3.4 GHz with a thermal design power (TDP) of 65W. I7-6700 is integrated into a desktop environment along with 8GB DDR4 RAM memory. However, many systems are designed in order to provide high performance with low power consumption. For that purpose, Intel has designed the Neural Compute Stick 2 (NCS2), a special platform that was based on Myriad X Visual Processor Unit. NCS2 can deliver high computational performance in computer vision tasks in about 1 watt. Both platforms have been tested under the Intel's OpenVino\textsuperscript{TM} toolkit, which enables deep learning inference at the edge and at the same time maximizes the performance of Intel's platforms. The preferred precision for DNN models is single-precision floating point (FP32) for CPUs, but other types of precision can be also supported, such as half-precision floating point (FP16) and fixed point. On the other hand, NCS2 can support only models with FP16 precision. 

\subsection{Nvidia}

High end GPUs are constantly used for achieving high throughput in computer vision tasks, either for training or inference. Such platforms are massively parallel and throughput oriented, but also they are a more power costly option compare to CPUs. An example is the mid-range GPU Nvidia GeForce GTX 1060 6GB, its peak power is 120W. GeForce GTX 1060 6GB integrates 1280 CUDA cores, but the absence of Tensor cores negatively affects the performance of FP16 precision. A power efficient option that is recommended by Nvidia is Jetson TX2. This embedded system is the second installment of the Jetson family and it is a 15W AI supercomputer at its maximum performance. Jetson TX2 integrates a 256-core GPU under Nvidia Pascal\textsuperscript{TM} Architecture, which supports the execution of models with both FP32 and FP16 precision by providing full-rate performance, contrary to the aforementioned platform. Furthermore, Jetson TX2 integrates a 8GB LPDDR4 RAM which is shared between the integrated GPU and CPU. In order to achieve higher performance on both hardware platforms provided by Nvidia, we have developed the computer vision tasks by implementing TensorRT. TensorRT is a special SDK, for increasing throughput and reducing latency in deep learning inference applications. In order for a platform to exploit the benefits of TensorRT, a system, that integrates a Nvidia's GPU, must support both CUDA and cuDNN.

\subsection{Arm}

Many power-constrained systems (i.e. mobile devices) integrate ARM Cortex-A series processors due to their relative high performance and very low power consumption, and, more recently, their support for 64-bit instruction sets. In order to fully utilise Cortex-A CPUs for deep learning inference, ARM provides the NN SDK that optimises the mapping of machine learning workloads on ARM devices. This SDK is accompanied by Arm's Compute Library, a repository of low-level functions for the accelerated execution of -mainly of computer vision- algorithms and functions on ARM processors. In our performance analysis we use the ARM Cortex-A57, a processor unit with 4 physical cores and a frequency of 2GHz, and the ARM Cortex-A53, a processor that integrates 4 physical cores clocked at 1.5 GHz. The utilised precision of the tested DNN models is FP32, but fixed point is also supported by the SDK. ARM Cortex-A57 is accompanied by 8 GB RAM, where ARM Cortex-A53 by 4 GB.  

\begin{table*}[t]
\centering
\caption{Inference Time (msec) per DNN model and platform (Batch Size = 1)}
\vspace{-0.05in}
\centering
\begin{tabular}{lrrrrrrrrr}
\hline
                & \textbf{Intel}   & \textbf{Intel} & \textbf{Nvidia}  & \textbf{Nvidia}           & \textbf{Nvidia}           & \textbf{Arm}        & \textbf{Arm}        & \textbf{Xilinx} & \textbf{Xilinx} \\
\textbf{Models} & \textbf{i7 6700} & \textbf{NCS2}  & \textbf{GTX 1060} & \textbf{TX2(FP32)} & \textbf{TX2(FP16)} & \textbf{Cortex-A57} & \textbf{Cortex-A53} & \textbf{ZCU102(DPU)} & \textbf{Alveo U50(DPU)}   \\ \hline
\textbf{Alexnet}       & 18.32    & 24.90   & 2.78             & 11.43            & 7.40             & 141.00  & 95.00  & -  & -   \\
\textbf{Inception-V1}  & 16.11    & 23.65   & 3.71             & 9.83             & 5.87             & 213.50  & 375.00  & 11.00 & 9.15\\
\textbf{Inception-V4}  & 94.69    & 141.98  & 22.34            & 83.92            & 41.41            & 1170.00 & 2078.50 & -   & 39.51  \\
\textbf{SqueezeNet1.1} & 3.13     & 10.21   & 2.15             & 3.38             & 2.32             & 77.50   & 106.00  & -   & 5.45  \\
\textbf{DenseNet 121}  & 27.00    & 50.20   & 13.37            & 29.74            & 20.21            & 339.00  & 667.00  & -    & - \\
\textbf{DenseNet 161}  & 66.53    & 139.38  & 22.58            & 66.03            & 45.46            & 773.00  & 1455.50 & -     & -\\
\textbf{DenseNet 169}  & 32.08    & 64.80   & 17.65            & 36.93            & 26.65            & 425.00  & 821.00  &       & -\\
\textbf{ResNet 50}     & 33.81    & 56.98   & 5.77             & 21.51            & 11.96            & 807.50  & 1008.50 & 21.50 & 14.74\\
\textbf{ResNet 101}    & 63.19    & 102.30  & 9.62             & 38.2             & 20.92            & 1441.00 & 1890.50 & 35.00 & 22.70\\
\textbf{ResNet 152}    & 92.97    & 152.95  & 13.98            & 55.15            & 30.18            & 2028.50 & 2662.50 & 48.00 & 30.96\\
\textbf{VGG16}         & 142.58   & 177.89  & 10.83            & 65.88            & 38.03            & 1029.50 & 1595.50 & 54.50 & 43.47\\
\textbf{VGG19}         & 145.24   & 215.96  & 12.70            & 79.32            & 45.66            & 1344.00 & 2026.50 & 62.50 & 50.52\\
\textbf{MobileNetV1}   & 5.58     & 20.67   & 1.91             & 6.09             &	5.36             & 82.20   & 158.10  & -     & -\\
\textbf{MobileNetV2}   & 5.51     &	31.16	& 3.16             & 8.79             &	7.49             & 127.60  & 214.80  & -	 & -\\
\hline
\textbf{SSD300}        & 184.40   & 610.55  & 21.12            & 118.17           & 117.64           & -       & -       & -     & -\\
\textbf{SSD512}        & 559.33   & 1383.74 & 43.39            & 279.07           & 278.85           & -       & -       & -     & -\\
\textbf{YOLO-V3}       & 204.25   & 596.23  & 36.85            & 277.00           & 253.93           & -       & -       & -     & -\\ \hline
\textbf{FCN8}          & 1117.67  & -       & 91.29            & 609.02           & 633.60           & -       & -       & -     & -\\
\textbf{Dilation}      & 14377.20 & -       & 1289.65          & 10738.42         & 10534.96         & -       & -       & -     & -\\ \hline
\end{tabular}
\label{InferenceResults}
\vspace{-0.15in}
\end{table*}

\subsection{Xilinx}


FPGAs are re-programmable platforms, that can be tailored to implement custom processor units or co-processors for specific tasks. Following this approach, Xilinx has released Vitis AI\footnote{https://github.com/Xilinx/Vitis-AI}, a development kit with all the comprehensive tools that enables the fast development of systems targeting FPGAs, providing at the same time a highly optimised mapping of a DNN workload to an FPGA. Along with the tools, Xilinx has designed a special Deep-Learning Processor Unit (DPU) that can be mapped into specific Xilinx FPGA platforms, in order to enable deep learning inference. However, DPU can only manage fixed point DNN models. Even though, the use of fixed point computations has important performance and resource advantages, the conversion is performed without losing prediction accuracy. The conversion of models into their fixed point precision as well as all DPU supported system calls for inference execution are part of Vitis AI. In our case, two boards are used for mapping the DPU; ZCU102 by using DNNDK and Alveo U50 by using VART. All model transformations into the fixed point precision are performed using the Vitis AI toolset. Furthermore, multiple threads can be used to increase the throughput and the overall performance of the DPU. The reported performance in the next section is the best we achieved by using multiple threads.

\begin{table}
\centering
\caption{Targeted platforms and SDKs}
\vspace{-0.05in}
\setlength\tabcolsep{2.5pt} 
\begin{tabular}{lll}
\hline
\textbf{Vendor} & \textbf{Platform} & \textbf{SDK}
\\ \hline
Intel  & Neural Compute Stick 2 (NC2) & OpenVino 
\\  \hline
Nvidia &  Jetson TX2 & TensorRT 
\\  \hline
Arm    & Cortex-A57 & ArmNN SDK 
\\  \hline
Arm    & Cortex-A53 & ArmNN SDK 
\\  \hline
Xilinx & ZCU102 (DPU) & DNNDK (Vitis AI) 
\\  \hline
Xilinx & Alveo U50 (DPU) & VART (Vitis AI) 
\\  \hline
Intel & i7 6700 & OpenVino 
\\  \hline
Nvidia & GeForce GTX1060 6GB & TensorRT 
\\  \hline
\end{tabular}
\label{Platforms}
\vspace{-0.15in}
\end{table}

\section{Evaluation}

The workloads (i.e. DNN models) defined in \autoref{DNNModels} are used for the evaluation of the above systems (Section \ref{HW}). Each workload was developed by integrating the corresponding SDK of the target platform with the main application. In order to provide an insight on the relative performance of the embedded systems with respect to systems without such constraints, the process was extended to target a desktop-rated CPU (Intel i7 6700 with 4 physical cores and 8 logical cores (using HyperThreading) with a base frequency of 3.4 GHz), and GPU (Nvidia GeForce GTX 1060 6GB, with 1280 CUDA cores and a peak power of 120W). \autoref{Platforms} lists the targeted platforms and their associated SDKs.

\subsection{System preparation}

Tensorflow and Caffe, two of the most widely used frameworks for machine learning application development, are supported by all platforms under investigation. The provided SDKs, accept models described in these frameworks and, prior to deployment, they perform their conversion to an Intermediate Representation (IR) format, performing a number of optimisations tailored to the targeted device architecture in order to improve the performance of the deployed system. By using the function calls that each toolkit provides, the new representation is mapped into the target platform for an optimised execution.    


In all DNN deployments that are investigated in this work, the above pre-processing step is performed capturing the impact of the tools' optimisations in the performance of the system. The reported inference time assumes that the data to be processed are already loaded on the off-chip memory of the system, and only reflects the time that it takes for the platform to execute the workload. More specifically, in the case of Jetson, Arm board, and Xilinx DPU, all the input data and parameters of the models are assumed to be on the memory of the board, where for both the i7 CPU and NCS2 the data are assumed to be on the host memory, while in GTX 1060 on the GPU board respectively.



It should be noted that the maximum batch size that could be investigated for each embedded platform was dictated by the amount of off-chip memory. As such, the maximum batch size is set to 128 for Image Classification, and to 8 for Image Segmentation. Nevertheless, a number of platforms failed to execute the workload under certain batch size values, as they were running out of memory.


\begin{figure*}[t]
    \vspace{-0.05in}
    \centering
    \includegraphics[width=1\textwidth]{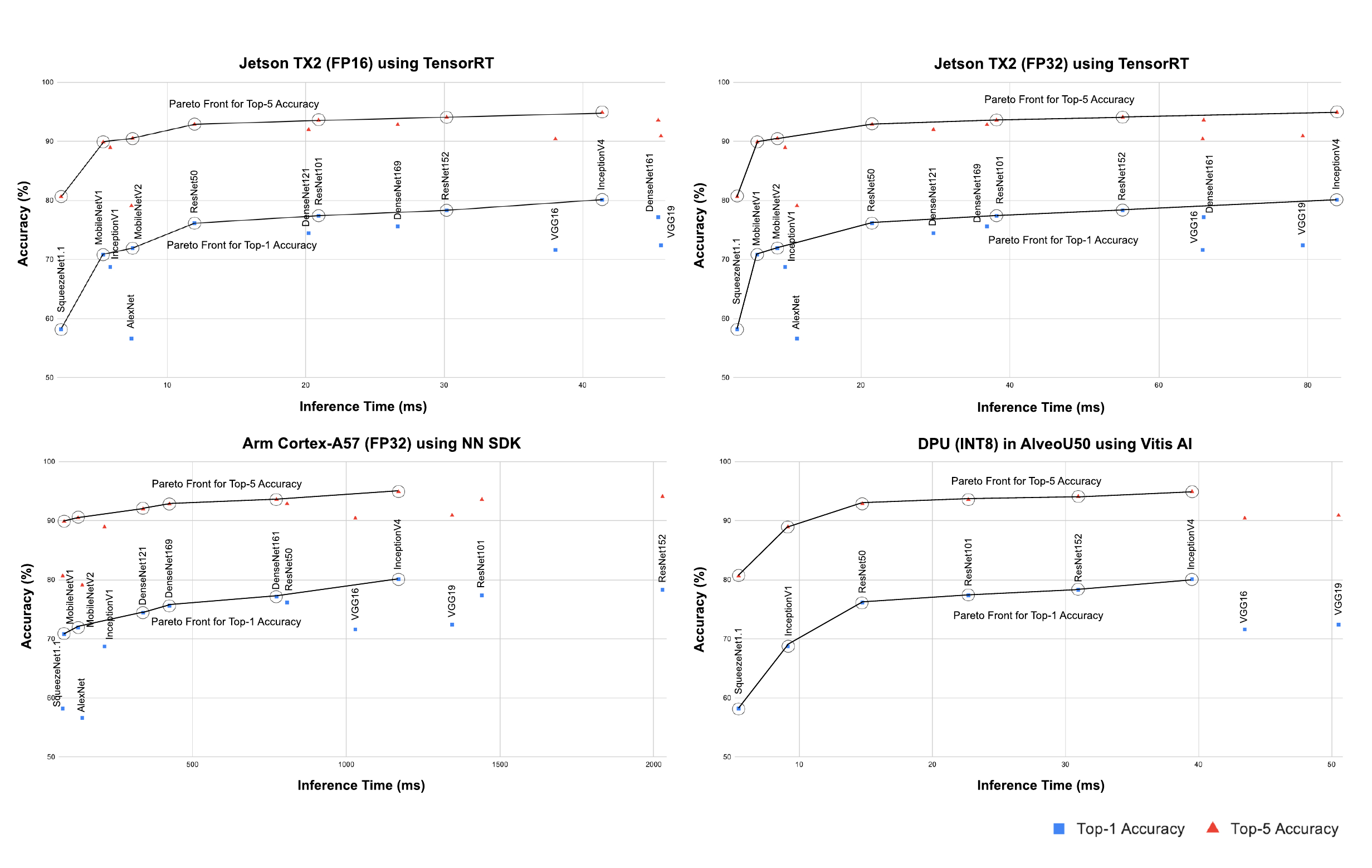}
        \vspace{-0.45in}
      \caption{ {Accuracy (\%) vs Inference Time (ms) and Pareto Front (black line) for Top-5 (red triangle) \& Top-1 (blue square)  accuracy for Image Classification Networks. At all cases Batch Size equals to 1.}}
  \label{fig:pareto}
      \vspace{-0.15in}
\end{figure*}




\subsection{Inference Time} \label{Inference Time}

\autoref{InferenceResults} reports the inference time of each platform-DNN model pair when a latency critical application is targeted, by setting batch size equal to 1.

Both Arm Cortex-A CPUs need considerably more time to execute the DNNs compared to all the other platforms, where the A57 achieves shorter inference time than A53, ranging between 24\% for ResNet152 and 49\% for DenseNet121. The next performant platform is NCS2, which demonstrated up to $18x$ faster inference results relative to the weak ARM processor for the Image Classification models, and its performance is comparable to the desktop CPU (i.e. i7);  both systems utilise Intel's OpenVINO\textsuperscript{TM} toolkit. However, NCS2 does not perform in par to the CPU device for Object Detection tasks, achieving inference times around 3 times longer that the i7 CPU. Image Segmentation performance results could not be obtained for the NCS2, as the application was timed out by the toolkit after a long execution.

Jetson TX2 outperforms Intel's NC2 platform by a factor of $2.52x$ on average, able as well to execute the Image Segmentation tasks. When an FP16 regime is used, TX2 is outperformed by the desktop GPU GTX1060 by a factor of $2.20x$ on average on Image Classification. One of the key differences between GTX 1060 and Jetson TX2 is the memory system: data in TX2 are fetched from a LPDDR4 RAM, while in GTX1060 from a Video RAM that delivers higher memory bandwidth. For Object Detection and Image Segmentation the obtained latency gap between the two platforms increases further, and the GTX 1060 is faster by an order of magnitude compare to TX2. Considering different precisions, FP16 in Jetson TX2 is faster by $35\%$ on average compare to FP32 across all Image Classification models. However, this improvement does not apply in the other two tasks, where the obtained latency results for both precision formats are similar. 


Finally, the performance using the Xilinx DPU on both FPGAs boards, Alveo U50 and ZCU102, fall between the TX2 under FP32 and FP16. All deployed models are in fixed point precision, and so their complexity and computational demands are significantly lower compared to their floating point-based models. Additionally, Alveo U50 achieves on average $37\%$ better performance compared to ZCU102 on image classification models. Meanwhile, results for some networks couldn't be extracted, as the tested models aren't supported by the Vitis AI toolchain.

\begin{figure*}{}
    \vspace{-0.05in}
    \centering
    \includegraphics[width=0.9\textwidth]{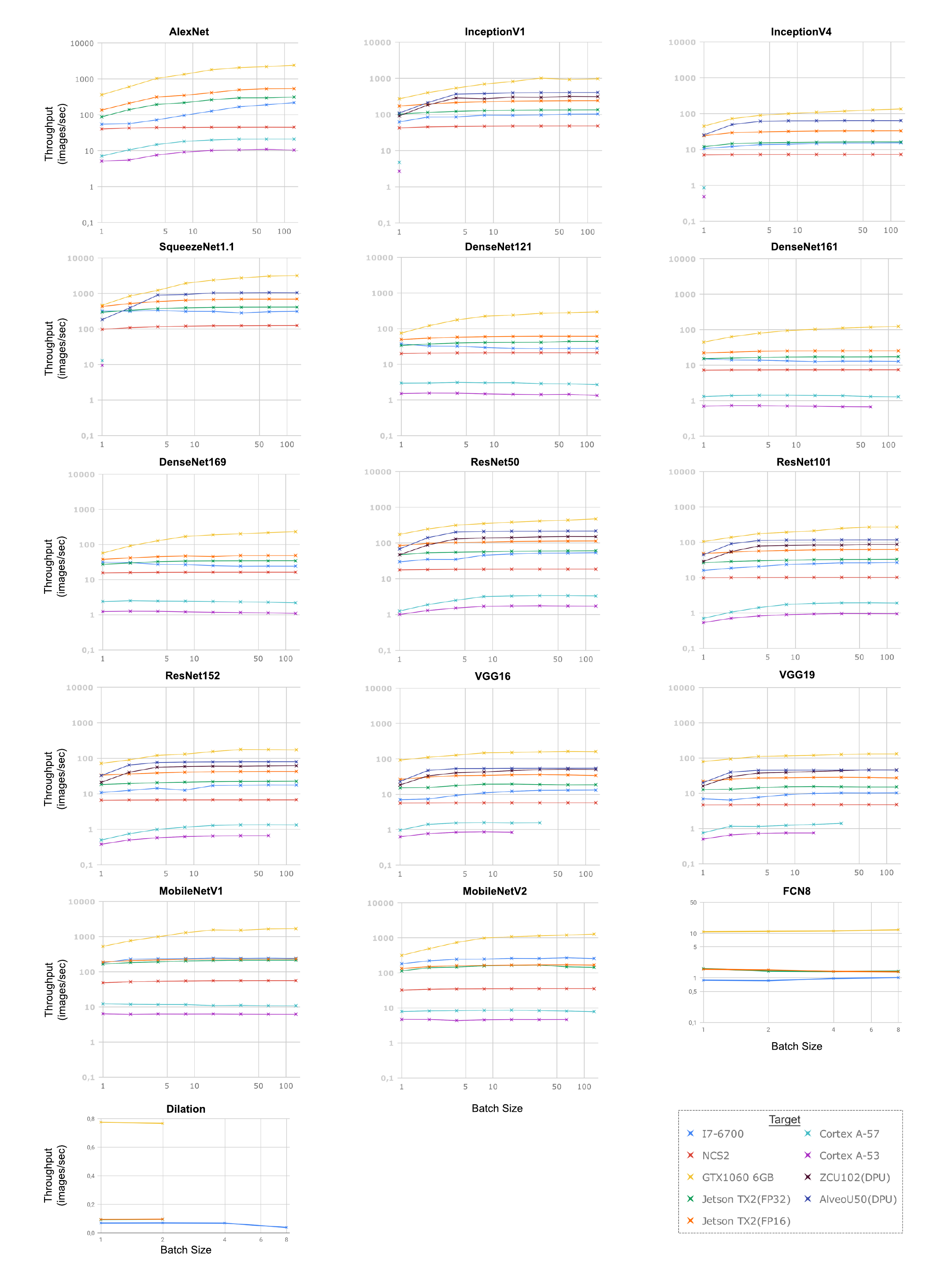}
  \vspace{-0.15in}
  \caption{Platform throughput for various DNN workloads and input batch sizes.}
  \label{fig:throughput}
  \vspace{-0.15in}
\end{figure*}

\begin{figure}[t]
    \includegraphics[width=\columnwidth]{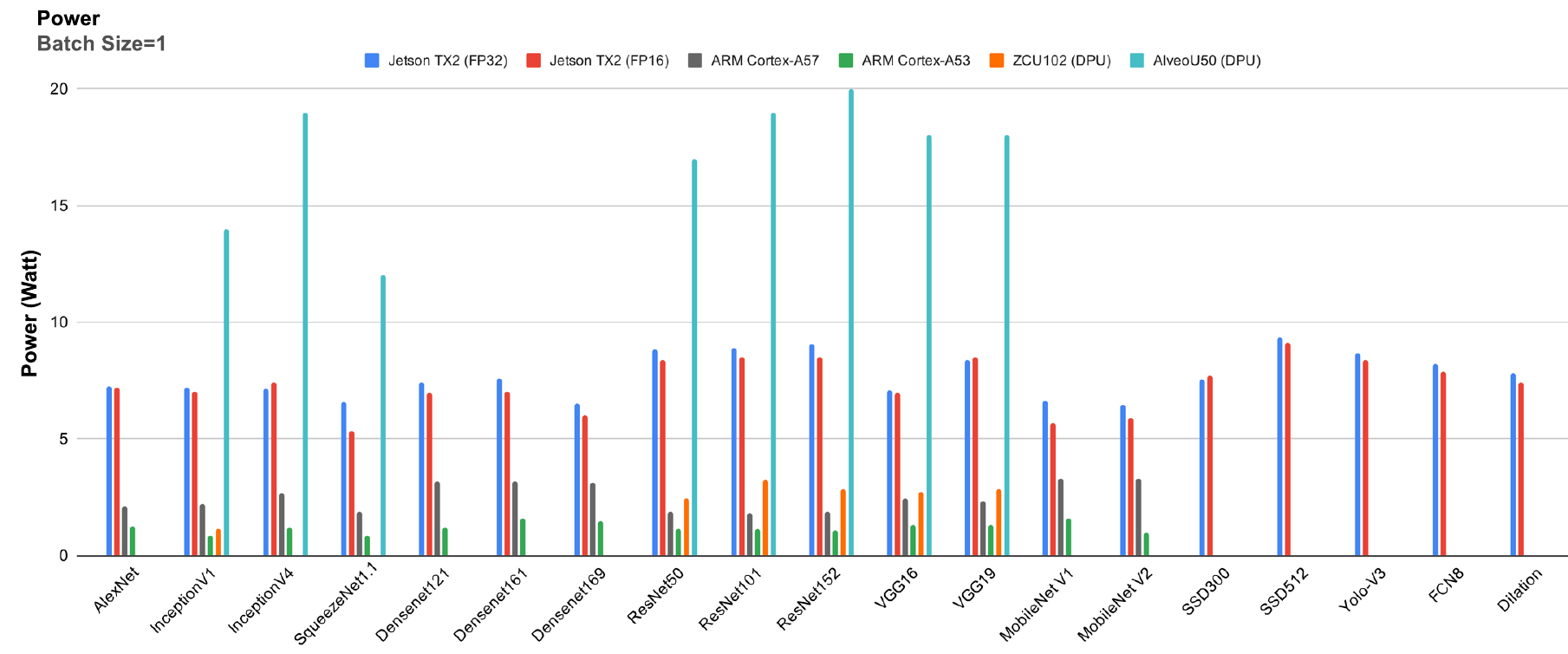}
  \caption{{Dynamic Power Consumption of embedded platforms and Alveo U50 for Batch Size = 1.}}
  \label{fig:power}
\end{figure}

\begin{figure}[t]
    \includegraphics[width=\columnwidth]{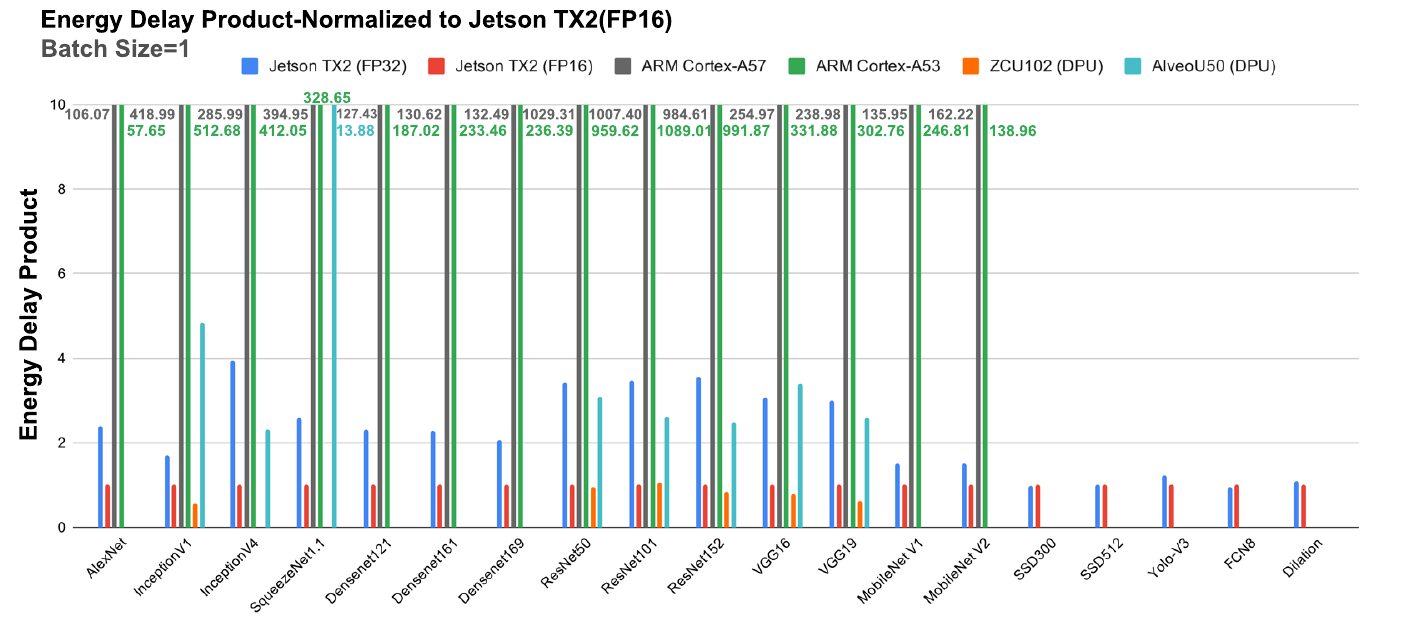}
  \caption{{Energy Delay Product of embedded platforms and Alveo U50 for Batch Size = 1.}}
  \label{fig:delay}
\end{figure}

\subsection{Pareto Efficiency}

While \autoref{InferenceResults} provides a clear image about the capabilities of each platform in various inference applications, a correlation between inference time and model's accuracy in a specific platform is needed. In \autoref{fig:pareto}, we have scattered the models from ImageNet dataset, based on Top-1 \& Top-5 Accuracy and inference Time. We select to visualize Jetson TX2 by tuning the application to both available representative formats, Arm Cortex A-57 tuning the application for FP32 models and the integrated DPU in Alveo U50 by using INT8 representative format, while keeping each batch to one single image for an inference run. Models with low inference time tend to have worse accuracy, as happens with SqueezeNet and AlexNet models. On the other hand InceptionV4 is one of the slowest tested models, however it provides the best results on both Top-1 \& Top-5 accuracy. 

Pareto Front tries to answer the question of which model is best for each case depending on the preferred platform. The previously described configurations include a large spectre across platforms and representative formats that can be leveraged by a developer. In Jetson TX2, in both representative formats, ResNet models tend to have higher accuracy and lower execution times than the DenseNet counterparts. Meanwhile, for applications where inference time is critical, the use of SqueezeNet or MobileNet models is preferred over AlexNet and InceptionV1, as the accuracy is significantly higher on each case. However, the benefits in performance utilizing FP16 precision apply differently in each network, Jetson TX2 columns in \autoref{Inference Time}. It is safely to assume that networks belonging in Pareto Front may vary by changing the network's precision in a single platform. On the other hand, Arm Cortex A-57 have quite different results. Densenet networks are preferred in the specific processor over the ResNet counterparts, while MobilenetV1 performs way better than Squeezenet1.1 by having the same inference time. Lastly, DPU on Alveo U50 tends to have similar behaviour with Jetson TX2, but the unavailability of few models reduces the choices. Generally, the use of a model that belongs to Pareto Front line is the appropriate choice depending on the requirements of the application and the targeted platform.

\subsection{Throughput}

In subsection \ref{Inference Time}, we evaluated the performance of our platforms with each batch including only one image during an inference run, and the comparison metric was the total inference time of the procedure. As platforms may perform better by processing multiple images in parallel, we evaluate the performance for batch sizes ranging from 1 to 128 for Image Classification and to 8 for Image Segmentation for all tested platforms. Then, we have calculated the throughput results for each scenario and report them in \autoref{fig:throughput}. Both axis in the graphs are in logarithmic scale in order to visualize better the differences in throughput between platforms.

\begin{figure}[t]
    \includegraphics[width=\columnwidth]{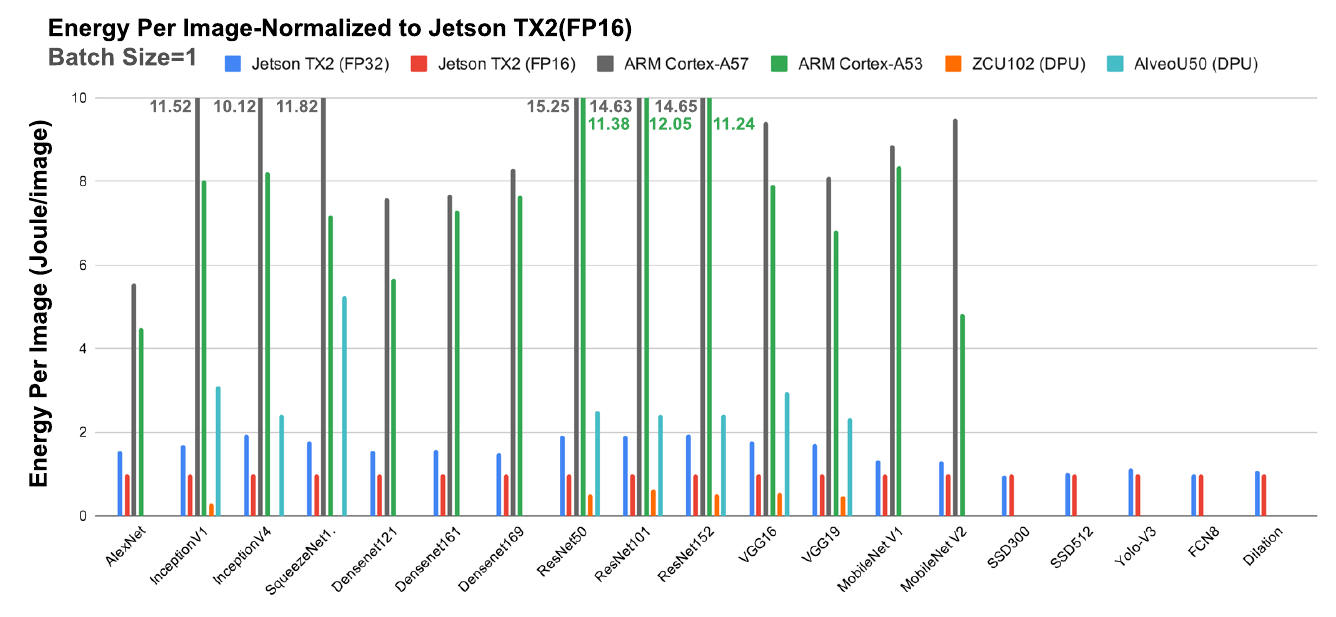}
  \caption{{Energy per Image of embedded platforms and Alveo U50 for Batch Size = 1.}}
  \label{fig:energy}
\end{figure}

The throughput of NCS2 has a minimal improvement that cannot even be clearly discerned in the graphs, irrespective of the DNN model and in comparison to the other platforms. The best performance of this platform is mostly  achieved by streaming batches with only one image, as inference time increases linearly with respect to batch size. On the other hand the two ARM Cortex-A CPUs exhibit the same behaviour as the batch size increases, again irrespective  of the model. The throughput of both processor units increases to a point where the metric cannot be further improved. Both ARM CPUs tend to perform better with larger batch sizes, but their performance is still very limited compared to the other platforms. Note that in the case of DenseNet and MobileNet models throughput tends to get worse as the batch size increases. 

As batch size is increasing, the high-end CPU from Intel behaviour depends on the DNN model of our task. Generally the throughput is slightly improving up to a point where a batch includes 16 or 32 images, except for AlexNet where the throughput keeps increasing further. However, for DenseNet and SqueezeNet the best performance is achieved when batch size equals to 1.

For the remaining platforms, as batch size increases the throughput is also improves and more images can be processed each second. However, this improvement is minimal on Jetson TX2, regardless of the model's precision. On the other hand, the throughput of the standalone GPU is greatly improved in all tested models, as this metric tends to keep increasing even in large batch sizes. Last but not least, by increasing the batch size from one to two or four, DPU in both FPGA platforms displays a huge improvement at its performance and the throughput is increased significantly. However, increasing the batch size further offers minimal improvement in throughput. 

The DNNs that are used for image segmentation task require a huge amount of MAC operations, by at least an order of magnitude more that the other models of our list. Due to limited memory resources on most of our platforms, measurements for larger batch sizes were not possible. Using the Dilation network, the throughput tends to worse in both CPU and standalone GPU, while in Jetson the value remains almost stable. On the contrary, FCN8 tend to be performed slightly better. Lastly, from both of these graphs we can export the conclusion that throughput in image segmentation task converge in both floating precision formats in Jetson TX2.

\subsection{Power Consumption and Energy per Image}

While inference time and throughput provide useful insights about the performance on a selected configuration (preferred model and platform), the power consumption of a target system running a vision kernel is another important factor, especially in embedded devices. \autoref{fig:power} reports real-time results of power consumption that each embedded platform needs to process batches of one image for each model. Each entry of the graph is obtained by measuring the vision kernel’s dynamic power while excluding the static power required to power the rest of the platform. So, each value represent the extra power that the platform consumes from an initial idle state. Furthermore, in \autoref{fig:energy}  we provide energy per image results of each configuration, while \autoref{fig:delay} reports the energy delay product (EDP). Both figures are normalized, where Jetson TX2, tuned with FP16 networks, is used as base for comparison between different platforms.

At first glance, Jetson TX2 consumes on average $7.72$ Watts when models with FP32 precision are used. By using a the lower supported representative format, FP16 for Jetson, the average power consumption can be reduced almost by a trivial $4.58$\%. On contrast, the use of ARM processors for computer vision inference tasks, can minimize the power consumption to a few watts. Especially, less than two watts are needed for Cortex-A53 in order to execute all ImageNet models, for image classification kernel. Finally, the power results for the platform that integrates the DPU accelerator are quite different from all above cases. The power that the FPGAs need for the inference procedure increases as the complexity of the model increases. So, a less computational intensive network with few MAC operations like InceptionV1 or Squeezenet1.1 needs fewer watts than ResNet or VGG networks. Meanwhile, using DPU with ZCU102 can challenge both the power efficient ARM CPUs. On the other hand, Alveo U50 consumes on average $17.13$ and up to $20$ Watts, a non-trivial amount compared to embedded platforms. However, this higher dynamic power consumption is expected from Alveo, as it is a FPGA card designed for Data Centers. 

Although, the use of a lower precision does not affect significantly the power consumption, the energy that each image needs is reduced by almost 20\%, mainly because of the lower inference time. So, the use of a FP16 model is both time and energy efficient. On the other hand, by using the power efficient ARM CPUs, each inference task needs considerably more energy than the other platforms, that exceeds $5x$ in some cases. However, the energy consumption of the integrated DPU in ZCU102, is lower even from Jetson TX2. So, an inference application can leverage the efficiency of the ZCU102 platform in both power and energy, by outperforming at the same time all platforms except the dedicated devices; GTX1060 and Alveo U50. The latter platform has an expected high energy consumption for the inference workload, due to the high needed power. 

Energy delay product is used as a metric to balance the energy consumption with the delay that occurs by the execution of the model during the inference task. Specifically, as we report in \autoref{fig:delay}, EDP for both ARM processors is exceptionally high compared to other platforms, as their energy consumption is significantly affected by the inference time. Meanwhile, the energy consumption on both Jetson TX2 and ZCU102 is not affected at all by inference time, as the results can be compared with the corresponding energy per image measurements in \autoref{fig:energy}. Finally, energy delay product in Alveo U50 depends on the kernel's workload. Workloads with low inference time, such as SqueezeNet1.1 and InceptionV1, present higher EDP compared to the rest of the models, as the execution time does not balance well with the energy consumption. 

\section{Discussion}

For portability reasons, mapping inference tasks onto particular commodity hardware platforms typically does not involve the use the platform-specific SDKs and/or libraries. Unfortunately, this approach leads to sub-optimal mapping as the generic approaches and libraries cannot exploit all the platform compute resources as well as it is possible using the more customized tools and libraries.
Also, the use of special processor units such NCS2 and DPU is not possible. In order to avoid such problems and also maximize the performance of an inference process, the use of software support is the only solution. At first, the existing toolkits transform the initial model into an optimal IR, and then they map it into the platform for a more efficient execution. 

During the previous section we offered an extensive overview of the performance and energy consumption of the most commodity platforms. While most of the exported results for inference time are quite good -even in platforms with RISC architecture like ARM Cortex-A series and to a greater extent in NCS2, in most cases the inference process includes large sets of images. Increasing the batch size affects differently each platform: in most cases, larger batch size leads to an increase of the throughput, though this improvement is limited by the memory bandwidth and the computational resources. Jetson TX2 is an example of this statement, as its shared memory between the CPU and GPU as well its limited memory bandwidth, limits the performance of the target integrated GPU. Similarly, the data transfer into NCS2 is a high cost procedure as it is plugged in a USB port, while this platform has also limited computational capabilities compare to a general purpose processor unit. 

A lot of effort is made towards developing customized networks to improve the performance of inference tasks targeting specific platforms, like FPGAs or GPUs. While we avoid utilizing such networks on our results, the scope of this paper is a unified performance landscape for a fair comparison between the tested platforms utilizing their respective SDK. However, most of the adopted networks are used as a starting point, either for application or network development. ResNet networks and Yolo-V3 are widely adopted networks from the community, while MobileNet dominates in latency-critical applications. Furthermore, both VGG16 and VGG19 are integrated as a core part of larger networks for feature extraction, like SSD for object detection and Dilation for image segmentation networks. 

The standard representative format for deep neural networks is FP32, as such a high precision format is very important for high accuracy for the networks' results. Maintaining FP32 format for network's training is necessary, as the network must propagate its results back in order to acquire better accuracy by modifying the weights of each layer. Keeping such a high precision is optional for inference tasks. So hardware vendors, especially Xilinx, are pushing the developers to utilize low precision networks in their inference tasks. Lowering the precision format to FP16, or using techniques like quantization to acquire INT8 networks, come over with trade-offs that have to do with lower accuracy in network's results. However, the benefits are huge, as Jetson TX2 is significantly faster using FP16 precision in Image Classification task, and NCS2 can compete even a high-end CPU. DPU is another example that utilizing INT8 precision networks can help outperform most of platforms, while keeping the task power and energy efficient.

In addition to image classification, other computer vision tasks are being used extensively by the community. While more and more models are developed to support these kind of tasks, the computational and memory demands of these DNNs are very large. Still, the increase of the batch size or the transformation of the model into FP16 precision may not affect the overall performance, as it tends to be stable in any case.

\section{Conclusion}


In this work, we have performed an experimental evaluation of embedded DNN platforms using their respective SDKs and state of the art networks for image classification, object detection and image segmentation. The evaluation shows that both Xilinx DPU and Jetson TX2 achieve the best performance for latency critical applications across all embedded devices, with the Jetson being able to support the execution of more state-of-the-art models compared to Xilinx DPU. Moreover, the evaluation of the embedded systems for throughput oriented applications shows that their performance under those benchmarks is largely insensitive to batch size, and maximum utilisation of the device can be achieved with modest batch sizes. The device that demonstrated the highest throughput gains with respect to batch size is the Xilinx DPUs, as its performance improves significantly when the batch size is increased from 1 to 4. Furthermore, insights about power demands and energy consumption are given for each configuration, confirming that the utilization of lower precision and the use of special processor units comes with significant benefits. In summary, the results provide a reference guide for the expected performance as well the real-time power and energy consumption of the above widely used platforms under state of the art DNNs workloads that could guide researchers and practitioners in adopting a specific platform for a target system.

\bibliographystyle{plain}
\bibliography{refs}

\begin{thebibliography}{10}

\bibitem{Embench}
M{\'{a}}rio Almeida, Stefanos Laskaridis, Ilias Leontiadis, Stylianos~I.
  Venieris, and Nicholas~D. Lane.
\newblock Embench: Quantifying performance variations of deep neural networks
  across modern commodity devices.
\newblock {\em CoRR}, abs/1905.07346, 2019.

\bibitem{blott}
M.~{Blott}, N.~{Fraser}, G.~{Gambardella}, L.~{Halder}, J.~{Kath}, Z.~{Neveu},
  Y.~{Umuroglu}, A.~{Vasilciuc}, M.~{Leeser}, and L.~{Doyle}.
\newblock Evaluation of optimized cnns on heterogeneous accelerators using a
  novel benchmarking approach.
\newblock {\em IEEE Transactions on Computers}, pages 1--1, 2020.

\bibitem{res}
K.~{He}, X.~{Zhang}, S.~{Ren}, and J.~{Sun}.
\newblock Deep residual learning for image recognition.
\newblock In {\em 2016 IEEE Conference on Computer Vision and Pattern
  Recognition (CVPR)}, pages 770--778, June 2016.

\bibitem{CaffePresso}
Gopalakrishna Hegde, Siddhartha, and Nachiket Kapre.
\newblock Caffepresso: Accelerating convolutional networks on embedded socs.
\newblock {\em ACM Trans. Embed. Comput. Syst.}, 17(1), November 2017.

\bibitem{MobileNets}
Andrew~G. Howard, Menglong Zhu, Bo~Chen, Dmitry Kalenichenko, Weijun Wang,
  Tobias Weyand, Marco Andreetto, and Hartwig Adam.
\newblock Mobilenets: Efficient convolutional neural networks for mobile vision
  applications.
\newblock {\em CoRR}, abs/1704.04861, 2017.

\bibitem{dense}
G.~{Huang}, Z.~{Liu}, L.~v.~d. {Maaten}, and K.~Q. {Weinberger}.
\newblock Densely connected convolutional networks.
\newblock In {\em 2017 IEEE Conference on Computer Vision and Pattern
  Recognition (CVPR)}, pages 2261--2269, July 2017.

\bibitem{squeeze}
Forrest~N. Iandola, Matthew~W. Moskewicz, Khalid Ashraf, Song Han, William~J.
  Dally, and Kurt Keutzer.
\newblock Squeezenet: Alexnet-level accuracy with 50x fewer parameters and
  {\textless}1mb model size.
\newblock {\em CoRR}, abs/1602.07360, 2016.

\bibitem{AI}
Andrey Ignatov, Radu Timofte, William Chou, Ke~Wang, Max Wu, Tim Hartley, and
  Luc~Van Gool.
\newblock {AI} benchmark: Running deep neural networks on android smartphones.
\newblock {\em CoRR}, abs/1810.01109, 2018.

\bibitem{AlexNet}
Alex Krizhevsky, Ilya Sutskever, and Geoffrey~E. Hinton.
\newblock Imagenet classification with deep convolutional neural networks.
\newblock In {\em Proceedings of the 25th International Conference on Neural
  Information Processing Systems - Volume 1}, NIPS’12, page 1097–1105, Red
  Hook, NY, USA, 2012. Curran Associates Inc.

\bibitem{ssd}
Wei Liu, Dragomir Anguelov, Dumitru Erhan, Christian Szegedy, Scott Reed,
  Cheng-Yang Fu, and Alexander~C. Berg.
\newblock Ssd: Single shot multibox detector.
\newblock {\em Lecture Notes in Computer Science}, page 21–37, 2016.

\bibitem{fcn8}
J.~{Long}, E.~{Shelhamer}, and T.~{Darrell}.
\newblock Fully convolutional networks for semantic segmentation.
\newblock In {\em 2015 IEEE Conference on Computer Vision and Pattern
  Recognition (CVPR)}, pages 3431--3440, June 2015.

\bibitem{mlperf}
Vijay~Janapa Reddi, Christine Cheng, David Kanter, Peter Mattson, Guenther
  Schmuelling, Carole-Jean Wu, Brian Anderson, Maximilien Breughe, Mark
  Charlebois, William Chou, Ramesh Chukka, Cody Coleman, Sam Davis, Pan Deng,
  Greg Diamos, Jared Duke, Dave Fick, J.~Scott Gardner, Itay Hubara, Sachin
  Idgunji, Thomas~B. Jablin, Jeff Jiao, Tom~St. John, Pankaj Kanwar, David Lee,
  Jeffery Liao, Anton Lokhmotov, Francisco Massa, Peng Meng, Paulius
  Micikevicius, Colin Osborne, Gennady Pekhimenko, Arun Tejusve~Raghunath
  Rajan, Dilip Sequeira, Ashish Sirasao, Fei Sun, Hanlin Tang, Michael Thomson,
  Frank Wei, Ephrem Wu, Lingjie Xu, Koichi Yamada, Bing Yu, George Yuan, Aaron
  Zhong, Peizhao Zhang, and Yuchen Zhou.
\newblock Mlperf inference benchmark, 2019.

\bibitem{yolov3}
Joseph Redmon and Ali Farhadi.
\newblock Yolov3: An incremental improvement.
\newblock {\em CoRR}, abs/1804.02767, 2018.

\bibitem{vgg}
Karen Simonyan and Andrew Zisserman.
\newblock Very deep convolutional networks for large-scale image recognition.
\newblock {\em arXiv 1409.1556}, 09 2014.

\bibitem{IncV1}
C.~{Szegedy}, {Wei Liu}, {Yangqing Jia}, P.~{Sermanet}, S.~{Reed},
  D.~{Anguelov}, D.~{Erhan}, V.~{Vanhoucke}, and A.~{Rabinovich}.
\newblock Going deeper with convolutions.
\newblock In {\em 2015 IEEE Conference on Computer Vision and Pattern
  Recognition (CVPR)}, pages 1--9, June 2015.

\bibitem{IncV4}
Christian Szegedy, Sergey Ioffe, Vincent Vanhoucke, and Alexander~A. Alemi.
\newblock Inception-v4, inception-resnet and the impact of residual connections
  on learning.
\newblock In {\em Proceedings of the Thirty-First AAAI Conference on Artificial
  Intelligence}, AAAI’17, page 4278–4284. AAAI Press, 2017.

\bibitem{dilation}
Fisher Yu and Vladlen Koltun.
\newblock Multi-scale context aggregation by dilated convolutions.
\newblock In {\em International Conference on Learning Representations (ICLR)},
  May 2016.

\end{thebibliography}

\end{document}